\documentclass[runningheads]{llncs}

\usepackage[T1]{fontenc}
\usepackage[utf8]{inputenc}
\usepackage{graphicx}
\usepackage{booktabs}
\usepackage{amsmath}
\usepackage{amssymb}
\usepackage{multirow}
\usepackage{tikz}
\usetikzlibrary{arrows.meta,positioning,fit,backgrounds,calc,%
                shapes.geometric,decorations.pathreplacing}
\usepackage{listings}
\usepackage{xcolor}
\usepackage{capt-of}
\usepackage{url}
\usepackage[hidelinks]{hyperref}
\usepackage{cleveref}

\newcommand{\inputdiagramas}[2]{%
  \begingroup
  \let\bvisoriginallabel\label
  \renewcommand{\label}[1]{\bvisoriginallabel{#2}}%
  \input{#1}%
  \endgroup
}

\newcommand{\sysname}{BVI}
\newcommand{\ccall}{c}                      
\newcommand{\org}{o}                        
\newcommand{\scred}{s_{\mathrm{cred}}}      
\newcommand{\spath}{s_{\mathrm{path}}}      
\newcommand{\sint}{s_{\mathrm{int}}}        
\newcommand{\sbeh}{s_{\mathrm{beh}}}        
\newcommand{\Dig}{\mathsf{D}}               
\newcommand{\Root}{\mathsf{root}}           
\newcommand{\Ptr}{\mathsf{ptr}}             
\newcommand{\ber}{\mathrm{BER}}             
\newcommand{\Reg}{\mathcal{R}}              
\newcommand{\Did}{\mathsf{did}}
\newcommand{\Sig}{\mathsf{Sig}}

\newcommand{\Hash}{\mathsf{H}}
\newcommand{\sid}{\mathsf{sid}}
\newcommand{\chid}{\mathsf{chid}}

\begin{document}
\emergencystretch=2em

\title{\sysname{}: Lightweight, Data-Centric Blockchain-Based\\Verification
of Identity Claims}
\titlerunning{\sysname{}: Lightweight, Data-Centric Verification of Identity Claims}
\author{Harshith Pothapala, Gagandeep Singh, Samudi Amarasinghe, Priyanka Singh, Xue Li}
\institute{The University of Queensland}

\maketitle

\begin{abstract}
A person who answers an unexpected call claiming to come from a bank has
no way to check the claim. Australian text messaging has labelled a
message as unverified when the sender identifier is not registered since
1 July 2026, but a voice call still arrives with nothing behind it, and
voice cloning has removed the last cue recipients relied on. We present
\sysname{}, which answers one question for the recipient: did the calling
party prove it holds a credential a registered organisation issued for
this call? \sysname{} keeps the organisational record, its authorised
channels and its revocation state on a public ledger in a directly
queryable form, and puts all decision logic in the handset, which performs
eleven checks, pays no transaction fee and holds no full-chain state. We
define ten attack classes plus the case in which revocation cannot be
resolved, compare them in a 10-by-4 coverage matrix spanning BBCA, the
content digest, the detector and composed BVI, and run 27 automated
scenario tests against isolated Hardhat fixture states of the BVI contract;
all 27 pass. For media substitution and synthetic speech, the contract tests
validate commitment and gating properties rather than claiming detector
accuracy. The mandatory per-session anchor costs about 91.5k gas and is
invariant in call duration and hop count; each participating intermediary
additionally contributes one optional attestation write. In a
20{,}000-session-per-cell path simulation, at one quarter carrier
participation BVI detects 20.6 per cent of randomly located in-path
rewrites versus 0.1 per cent for the every-hop comparison, while identity
and content evidence remain available even when path evidence is
indeterminate. We also quantify the throughput ceiling of the anchoring
design.
\end{abstract}

\noindent\textbf{Anonymous artefact.}
The implementation and evaluation artefact is available at
\url{https://anonymous.4open.science/r/bvi-D415}.

\section{Introduction}
\label{sec:intro}

At 4:12 on a Tuesday afternoon a phone rings. The caller says she is from
the fraud team at the recipient's bank, that a payment has been flagged,
and that she needs to confirm a few details before the account is frozen.
The handset shows a number and, if the network happens to support it, a
name pulled from a directory. The recipient now has to decide. The only
evidence available is how the caller sounds and how well she knows the
script, and both of those are now cheap to manufacture.

We take this call as the running example of the paper and return to it in
every section. We write S0 for the honest version of it, A1 to A10 for
the ten ways the same call can be dishonest, and U for the case in which
the handset cannot resolve enough to say. Throughout we ask what the
handset could show the recipient at 4:12 that she could act on.

\subsection{The Asymmetry Between Text and Voice}

Australia has already answered a version of this question for text. The
\emph{Telecommunications (SMS Sender ID Register) Industry Standard
2025}, made under Part 24B of the \emph{Telecommunications Act 1997}
(Cth), requires providers to over-stamp any unregistered alphanumeric
sender identifier with the word ``Unverified'' and to group those
messages away from the registered ones. Registration opened on
30 November 2025 and over-stamping took effect on 1 July
2026~\cite{acma2025smsstd,acma2025smsrules}. A recipient reading a text
now gets a registry-backed statement about the sender before she reads a
word of it.

No equivalent exists for a voice call. The secure telephone identity
framework~\cite{rfc8224,rfc8225,rfc8226,shaken2022} signs a numbering
resource and attaches an attestation level, which is a statement about a
phone number and its assignment rather than about the organisation the
caller claims to represent, and it says nothing about what happens after
the call connects; Wang et al.\ document its practical limits, including
the difficulty of scaling its certificate infrastructure across
jurisdictions and its silence on legs not carried over
IP~\cite{wang2024spoofing}. Branded calling and the Australian
\emph{Reducing Scam Calls and Scam SMs} industry code~\cite{ca2022c661}
improve what appears on the screen within one operator's domain and stop
the moment the call is answered. The recipient in S0 is therefore worse
off on a voice call than on a text from the same organisation, which is
the wrong way round, because voice is the channel where an attacker can
improvise.

\subsection{Why This Matters Now}

The National Anti-Scam Centre reports combined losses of AUD 2.18 billion
across its reporting partners in 2025, up 7.8 per cent, from 481{,}523
reports of which 274{,}577 involved a loss~\cite{nasc2026}. The
regulatory response places duties on the organisations in the middle
rather than on the recipient. The \emph{Scams Prevention Framework Act
2025} (Cth) inserted Part IVF into the \emph{Competition and Consumer Act
2010} (Cth), imposing governance, prevention, detection, reporting,
disruption and response obligations with penalties reaching AUD 50
million per contravention~\cite{spfact2025}. Banking,
telecommunications and digital-platform services were designated by
instrument F2026L00627, registered on 28 May
2026~\cite{spfdesignation2026}. Treasury also released exposure-draft
rules and sector codes for consultation in May
2026~\cite{treasury2026codes}; the \emph{Competition and Consumer (Scams
Prevention Framework) Rules 2026} were subsequently made and registered
on 31 August 2026~\cite{spfrules2026}. Transitional provisions defer
most substantive sector obligations until 31 March 2027. Two things
follow. A bank must be able to
evidence after the fact that a call it made was the call it says it made,
and a recipient needs more than a brand name on a screen, because the
screen is what an attacker controls most easily.

\subsection{What Existing Work Leaves on the Table}

Four bodies of work bear on S0. Ledger-based caller identity registers an
organisation or a number against a tamper-evident
record~\cite{chen2021callchain,kara2023voipchain,liu2020authentic}, and
all of it finishes when the session is established. BBCA re-checks the
calling identifier at every hop until termination, so a transit element
that rewrites the origin becomes detectable~\cite{tas2024bbca}; but it
authenticates a number rather than an organisation, publishes no
implementation or measurement of any kind, defines no behaviour for a
carrier outside the ledger, and answers the privacy objection by
encrypting caller data on a record that cannot be deleted. Cryptographic
content authentication exchanges robust digests over a side channel and
separates legitimate network modification from
tampering~\cite{reaves2017authenticall,reaves2016authloop}, but its trust
root is a central service that also retains the result. Notably, BBCA
cites that work as caller identifier verification, omitting the content
authentication that is half its title: the ledger line has inherited the
identity half and left the content half behind. Learned detection sees
synthesis but does not transfer beyond its training
distribution~\cite{muller2022generalize,barrington2024deepspeak}, and
watermarking~\cite{sanroman2024audioseal} needs the generator to have
cooperated. None of this puts anything in the recipient's hand at 4:12
that is cheap enough for every call and specific enough to act on.

\subsection{What \sysname{} Does}

\sysname{} answers a narrower question than the literature usually asks,
and answers it completely: \emph{did the calling party prove, for this
call and no other, that it holds a credential issued by an organisation
whose ledger record is currently active and whose authorised channels
include the one the call arrived on?}

The handset asks for that proof, reads the organisation's record, runs
eleven checks, and returns a credential score $\scred$. It writes
nothing, pays no transaction fee, keeps no full-chain state, and performs
no call-time chain write. Three layers surround this one, covering the
route, the audio, and the residual case in which a registered key is in
the wrong hands.

Two words in the title are deliberate. \emph{Lightweight} means the
verifier carries no operational burden: no gas, no full node, no
subscription, and no call-time transaction write. \emph{Data-centric}
means the ledger holds the identity facts in a directly addressable
shape, so verification is a lookup over structured state rather than a
replay of an event log through an intermediary index. The second is what
makes the first achievable without reintroducing a trusted service.

\subsection{Contributions}

\textbf{C1.} An attack catalogue for organisational impersonation on
unsolicited calls: ten attack classes and one indeterminate state, each
stated in terms of what the recipient sees and each a distinct code path
in the verifier (\Cref{sec:scenarios}).
\textbf{C2.} A requirement-driven ledger selection over six candidates
against light-client verifiability, state shape, cost model and
permission model, resolved as a split deployment that keeps revocation
off a single-sequencer ordering path and places per-session anchors where
they can be afforded (\Cref{sec:ledger}).
\textbf{C3.} A four-layer design in which the organisation, not its
carrier, is the subject of the record, path attestation is optional and
session-bound, and the detector may only remove a verdict it never
granted, with a proposition establishing that gating is the only
non-degenerate linear composition (\Cref{sec:design}).
\textbf{C4.} A prototype of the registry, path attestation,
session anchoring, digest pipeline and composition gate, together with a
10-by-4 attack-coverage matrix and 27 automated Hardhat tests grouped by
A1--A10; all 27 pass under isolated Hardhat fixture states
(\Cref{sec:impl,sec:eval}).
\textbf{C5.} Evaluation (\Cref{sec:eval}): measured gas for the
on-chain operations, a roughly 91.5k-gas per-session anchor whose cost is
independent of duration and hop count, path coverage under partial
participation against average-case and strategic attacker models,
controlled splice detection at fixed false positive rates, and a derived
throughput ceiling.

We are explicit about what is not ours. The robust speech digest is due
to AuthentiCall~\cite{reaves2017authenticall} and the fingerprinting it
rests on~\cite{haitsma2002robust,cano2005review}. Hop-by-hop ledger
verification of a calling identifier is due to BBCA~\cite{tas2024bbca}.
We propose no new synthetic speech detector.

\section{The Problem from the Recipient's Side}
\label{sec:scenarios}

\subsection{Eleven Ways S0 Can Go}

Scenario S0 is the honest case: a real agent of a real bank, calling for
a real reason. \Cref{tab:attacks} sets out the ten ways the same call can
be dishonest, plus the case in which the handset cannot tell. We derive
these from the shape of the credential presentation and the shape of the
session rather than from a threat-modelling framework, because each is
something an implementer must write code for and each is what the
recipient would describe if asked what went wrong.

\begin{table}[tb]
\centering
\caption{The attack catalogue. Each row is a distinct thing that can be
wrong with the call at 4:12, a distinct code path in the verifier, and a
distinct test group in \Cref{sec:impl}. U is not an attack; it is the
state in which the handset cannot resolve revocation and must say so
rather than guess.}
\label{tab:attacks}
\footnotesize
\setlength{\tabcolsep}{4pt}
\begin{tabular}{@{}llp{4.6cm}l@{}}
\toprule
ID & Attack & What the calling party does & Layer \\
\midrule
A1  & No credential          & Presents nothing and relies on the calling identifier alone & L1 \\
A2  & Self-signed credential & Presents a credential it issued to itself & L1 \\
A3  & Stolen credential      & Presents a genuine credential issued to somebody else & L1 \\
A4  & Tampered claim         & Presents a genuine credential with a field edited & L1 \\
A5  & Revoked credential     & Presents a credential the issuer has since withdrawn & L1 \\
A6  & Expired key epoch      & Signs under a key whose validity window has closed & L1 \\
A7  & Replayed assertion     & Replays a capture of a genuine earlier presentation & L1 \\
A8  & In-path rewrite        & Rewrites the calling identifier at a transit element & L2 \\
A9  & Media substitution     & Replaces or splices audio after authentication & L3 \\
A10 & Synthetic speech       & Speaks in cloned voice under a valid credential & L4 \\
U   & Status unknown         & Presents a well-formed credential the handset cannot fully resolve & L1 \\
\bottomrule
\end{tabular}
\end{table}

A1 through A5 and A7 are adversarial and must fail closed. A6 differs in
kind: a revoked credential is usually a former employee or a compromised
key and should be treated as hostile, while an expired one is more often
a badge that timed out over a weekend, where the right answer is to
withhold the verified indicator without accusing the caller of anything.
U is the honest admission that if the revocation status cannot be
reached, the handset knows less than it would like, and saying so beats
guessing in either direction. This three-way distinction is the whole
reason $\scred$ is graded rather than binary.

A8, A9 and A10 are the reason the design has more than one layer, and
they are worth reading together. In A8 the audio is authentic and the
route is not, so a digest cannot see it. In A9 the route is authentic and
the audio is not, so a transcript cannot see it. In A10 both are
authentic and the speaker is a machine, so neither can see it. A scheme
carrying one mechanism misclassifies the attack it cannot see as no
attack at all.

\subsection{Requirements}

Six requirements follow from S0, from A1 to A10 and U, and from the
regulatory position above.
\textbf{R1, no prior trust:} verification succeeds with no shared secret,
prior contact or out-of-band channel, because the recipient in S0 has
never spoken to this agent.
\textbf{R2, no shared operator:} verification requires no shared carrier,
platform or vendor, and no account with any verification service, which
rules out a subscription trust root.
\textbf{R3, one session only:} a presentation is valid for exactly one
call on exactly one channel and cannot be transferred, or A7 succeeds.
\textbf{R4, no recipient transaction burden:} no transaction fee, no
full-node operation and no call-time chain write, because a defence that
requires the recipient to transact or operate infrastructure is one most
recipients will not have.
\textbf{R5, incremental deployability:} the guarantee degrades gracefully
in the number of participating intermediaries, and a call over a wholly
non-participating path is still verifiable at the identity layer, so no
precondition is placed on carriers before the scheme is useful to
anybody.
\textbf{R6, evidence after the call:} any party, including a regulator or
a court, can check a claim about what the endpoint asserted against a
record no participant in the dispute controls.

R4 is the requirement this literature most often skips and it decides
whether a design can reach a handset at all. R2 and R6 rule out a plain
signed registry file served by one party, but they do not logically force
a blockchain: an independently monitored append-only transparency log can
also expose rollback or equivocation~\cite{rfc9162}. The next section
states why we nevertheless choose a permissionless ledger for this threat
and governance model.

\section{Why a Ledger, and Which One}
\label{sec:ledger}

\subsection{The Signed Registry File Objection}

A distributed ledger audience should ask why a signed registry file
served over HTTPS does not solve this. It nearly does. Organisations are
listed, the file is signed, handsets fetch it. A plain file, however, can
be silently rolled back by its serving party and does not by itself give
an independently auditable statement of when a credential's status
changed. Append-only transparency logs are a genuine alternative:
Merkle consistency and inclusion proofs can expose rollback and some
forms of log misbehaviour~\cite{rfc9162}. They retain a log-operator trust
and monitoring model, including the need to detect inconsistent views.
Our design instead uses a permissionless ledger so the status transition
and its ordering are established by the ledger consensus rather than by a
single registry operator. The claim is therefore not that only a
blockchain can provide auditable history, but that this trust model
matches R2 and R6 without introducing a separate log operator. The
per-session anchor serves the same evidential goal for session data.

\subsection{Selection Criteria}

Five criteria follow, four from the requirements and one from the design.
\emph{Permissionless read and light verification}, because R2 requires a
verifier to check a read against a block header without a full node and
without trusting an operator's endpoint, which would reintroduce the
trust root we removed. \emph{State shape}, because verification is a
handful of point lookups, and a platform whose queryable state is a
transaction log requiring an off-chain indexer puts that indexer on the
trust path and costs us R2 again; this is what we mean by data-centric,
and it is as much a property of our storage layout as of the platform.
\emph{Cost model}, because recipients pay nothing (R4) while
organisations pay for registration, amortised over call volume.
\emph{Contract language and toolchain maturity}, because a registry is
long-lived infrastructure and migration is a governance event.
\emph{Permission model}, because registration is permissioned while
verification is not, so a platform coupling the two forces us to give up
one.

The platform comparison follows current platform documentation and distinguishes
consensus openness from public readability. Ethereum provides protocol-level state-root verification;
the standard OP Stack model has a sequencer plus an L1 forced-inclusion
path; Solana exposes standard account reads through RPC; Fabric and
Corda use membership-controlled networks; and CometBFT leaves
application-state proof semantics to the application
layer~\cite{ethereumLight2026,optimismForced2026,solanaCore2026,fabricGlossary2026,cordaNetwork2026,cometbftABCI2026}.

\subsection{The Choice, and What It Costs Us}

We deploy the registry as an Ethereum contract~\cite{wood2014yellow,ethereumLight2026}. Permissioned platforms such as Fabric and Corda put network access under
membership control and therefore do not meet R2 for an unaffiliated
recipient. Solana's account model is attractively data-centric, but the
standard RPC read path we evaluate does not provide the proof-carrying
state read R2 requires.

That leaves a decision between the base layer and a rollup, and the two
classes of write in \sysname{} want opposite answers. Revocation is the
operation where delay is a security property rather than an
inconvenience. A default OP Stack deployment has a single sequencer,
although users can bypass it through L1 forced
transactions~\cite{optimismForced2026}; this still introduces a distinct
sequencing-window and finality model. Per-session anchoring is the
operation where recurring cost decides whether the scheme can run at all.
Because spot fee prices vary, our evaluation compares measured gas and
write counts rather than treating a dollar-per-call estimate as a stable
system property.

We therefore split the deployment rather than pretending one platform
dominates. The registry and every status transition, including
revocation, are written to Ethereum L1, so ordinary inclusion does not
depend on a single rollup sequencer. This is not a claim of instantaneous
or censorship-proof inclusion: L1 still has block production and
finality delay. Per-session anchors are written to an OP Stack rollup,
where the recurring write can be provisioned separately. The registry read a handset performs at
4:12 is an L1 read and carries the revocation guarantee; the anchor it
may later be asked about is a rollup read and carries the evidential
guarantee, and the two are bound together by the session identifier
rather than by the platform. The cost of the split is that a dispute
resolved months later involves two chains, and that rollup settlement
delay bounds how quickly an anchor becomes final. We state the throughput
consequence in \Cref{sec:eval:throughput} rather than leave it to be
found.

Ethereum gives us no native query language, so we recover the
data-centric property through storage layout instead: records in a
mapping keyed by the hash of the identifier, authorised channels in a
nested mapping so membership is a single storage read rather than a scan,
attestations keyed by session and hop, anchors keyed by organisation and
session. Every read a verifier needs is one indexed lookup, which is what
keeps the recipient's work at the level \Cref{sec:eval} measures.

\section{Design}
\label{sec:design}

\subsection{Four Layers over One Ledger}

\begin{figure}[t]
\centering
\resizebox{0.98\linewidth}{!}{%
\begin{tikzpicture}[
  font=\scriptsize,
  box/.style={draw,rounded corners=2pt,align=center,inner sep=4pt,minimum height=8mm},
  chain/.style={box,fill=blue!8,text width=27mm},
  layer/.style={box,fill=blue!5,text width=37mm},
  behaviour/.style={box,fill=orange!12,text width=37mm},
  endpoint/.style={box,fill=gray!7,text width=22mm},
  decision/.style={box,text width=18mm,minimum height=10mm},
  arr/.style={-{Latex[length=1.8mm]},thick},
  darr/.style={-{Latex[length=1.8mm]},thick,dashed},
  note/.style={font=\tiny,fill=white,inner sep=1.5pt,align=center}
]

\node[chain] (attest) at (4.40,5.15)
  {Attestation store\\[-1pt]$\mathsf{attest}(\sid,j,\cdot)$};
\node[chain] (registry) at (7.40,5.15)
  {Registry $\Reg$\\[-1pt]$(\Did_\org,pk_\org,\mathcal{C}_\org,\mathsf{status})$};
\node[chain] (anchors) at (11.65,5.15)
  {Anchor store\\[-1pt]$(\Root_\ccall,\Ptr_\ccall)$};
\begin{scope}[on background layer]
\node[draw,rounded corners=3pt,fill=blue!2,fit=(attest)(registry)(anchors),
      inner xsep=5mm,inner ysep=5mm] (chainband) {};
\end{scope}
\node[font=\scriptsize\bfseries,anchor=south,fill=white,inner sep=1pt]
  at ([yshift=1mm]chainband.north) {On-chain state};

\node[endpoint] (caller) at (0.90,1.35) {Calling party $P$\\claims $\org$};
\node[endpoint,text width=24mm] (path) at (4.40,1.35)
  {Path $R_1,\ldots,R_h$\\subseteq $Q$ attests};
\draw[arr] (caller.east) -- node[above,note] {INVITE} (path.west);

\node[layer] (l1) at (7.40,2.70)
  {\textbf{L1 Identity}\\[-1pt]$\sigma$ over $(n,\chid,t,k_\ccall)$};
\node[layer] (l2) at (7.40,1.55)
  {\textbf{L2 Path}\\[-1pt]transcript $\mathcal{T}_\ccall,\ \spath$};
\node[layer] (l3) at (7.40,0.40)
  {\textbf{L3 Content}\\[-1pt]$d_i=\Dig(a_i),\quad \ber\leq\tau_{\ber}$};
\node[behaviour] (l4) at (7.40,-0.75)
  {\textbf{L4 Liveness}\\[-1pt]$\sbeh$, demote only};
\draw[arr] (l1.south) -- (l2.north);
\draw[arr] (l2.south) -- (l3.north);
\draw[arr] (l3.south) -- (l4.north);

\node[decision,fill=green!11] (gate) at (10.45,0.98)
  {\textbf{Gate}\\Eq.~\ref{eq:gate}};
\node[decision,fill=green!18] (verified) at (13.05,1.70) {Verified};
\node[decision,fill=red!10] (notverified) at (13.05,0.20) {Not\\verified};
\coordinate (busTop) at (9.38,2.70);
\coordinate (busBottom) at (9.38,-0.75);
\draw[thick] (busTop) -- (busBottom);
\draw[thick] (l1.east) -- (l1.east -| busTop);
\draw[thick] (l2.east) -- (l2.east -| busTop);
\draw[thick] (l3.east) -- (l3.east -| busTop);
\draw[thick] (l4.east) -- (l4.east -| busTop);
\draw[arr] (9.38,0.98) -- (gate.west);
\draw[arr] (gate.east) -- ++(0.40,0) |- (verified.west);
\draw[arr] (gate.east) -- ++(0.40,0) |- (notverified.west);

\draw[arr] (path.east) -- ++(0.30,0) |- (l1.west);
\node[note] at (5.35,2.18) {claim + session setup};
\draw[arr] (path.east) -- ++(0.55,0) |- (l3.west);
\node[note] at (5.25,0.06) {audio $+\ d_i$};

\draw[darr] (registry.south) -- (l1.north);
\draw[darr] (path.north) -- (attest.south);
\draw[darr] (attest.east) -- ++(0.45,0) |- ([xshift=-2mm]l2.west);

\begin{scope}[on background layer]
\node[draw,dashed,rounded corners=4pt,fit=(l1)(l2)(l3)(l4)(gate),
      inner xsep=5mm,inner ysep=6mm] (recipient) {};
\end{scope}
\node[font=\scriptsize\bfseries,anchor=north west,fill=white,inner sep=1.5pt]
  at ([xshift=3mm,yshift=-1mm]recipient.north west) {Recipient verifier};

\draw[darr] (caller.south) -- ++(0,-3.55) -- ++(13.55,0)
  -- ++(0,7.35) -- (anchors.east);
\node[note] at (10.55,-2.28) {one anchor at teardown};
\end{tikzpicture}}
\caption{\sysname{} architecture. The recipient evaluates four layers before
presenting a verified indicator. L1 establishes the organisation and
session-bound credential, L2 evaluates optional session-bound path evidence,
L3 checks media integrity, and L4 supplies a statistical liveness signal that
may only demote a verdict. Audio and digests remain inside the recipient trust
boundary. The ledger stores organisational state, optional hop attestations,
and one joint session anchor. The anchor write is constant in call duration and
hop count, while optional path-attestation writes scale with the number of
participating intermediaries.}
\label{fig:arch}
\end{figure}
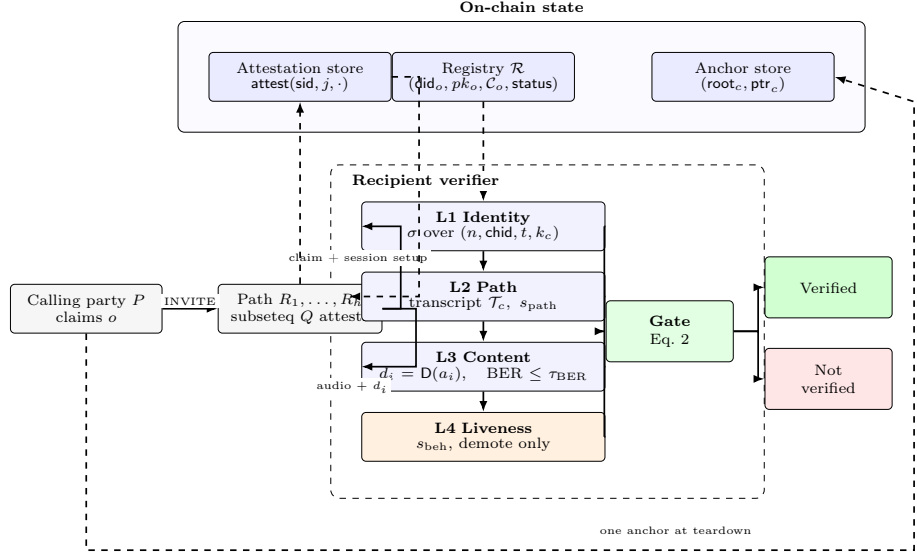

\Cref{fig:arch} shows the arrangement. The identity layer holds
organisational records and authorised channels on chain and issues a
per-session assertion bound to a recipient nonce and to the arrival
channel. The path layer collects optional attestations from participating
intermediaries, each bound to that assertion, and reduces them to an
ordered transcript. The content layer computes robust digests at both
endpoints, compares them during the call, and commits to the digest
sequence and the transcript together in one write at teardown. The
liveness layer runs a learned detector locally and may only demote. The
ordering is not decorative: the first three layers are cryptographic and
carry the guarantee, while the fourth is statistical and is confined,
because its error rate against a generator an adversary chooses is high
and cannot be known in advance.

\subsection{Layer 1: The Organisational Record and the Eleven Checks}
\label{sec:design:l1}

Each participating organisation publishes a record
\[ \Reg[\org] = (\Did_\org,\ pk_\org,\ \mathcal{C}_\org,\ m_\org,\
\mathsf{status}_\org), \]
in which $\Did_\org$ is a decentralised identifier following the W3C data
model~\cite{w3cdid2022}, $pk_\org$ the current verification key,
$\mathcal{C}_\org$ the authorised channel identifiers, $m_\org$ a
content-addressed pointer to off-chain metadata such as legal name and
business number, and the status is active, suspended or revoked.
Credentials issued to outbound platforms follow the verifiable credential
data model~\cite{w3cvc2025}. Channels are stored as hashes rather than
plaintext identifiers, so a verifier that already observes an identifier
can test membership without publishing the literal value in contract
state. This is pseudonymisation, not strong secrecy: low-entropy spaces
such as telephone numbers remain susceptible to dictionary enumeration.

This is where we depart from BBCA, where an enterprise is reachable only
through its provider's registration, so the anchor is the set of
participating providers and changing carrier changes
anchor~\cite{tas2024bbca}. Here the organisation is the subject and the
channel an attribute of it, so the anchor is independent of who carries
the call, which is what R2 demands.

During setup, for a call $\ccall$ claiming $\org$: the calling party
signals the claim carrying $\Did_\org$; the handset draws a nonce
$n \leftarrow \{0,1\}^{128}$ and returns it; the calling party replies
with $\sigma = \Sig_{sk_\org}(\Hash(\Did_\org \Vert n \Vert \chid \Vert t
\Vert k_\ccall))$, where $\chid$ is the hashed identifier of the channel
carrying the call, $t$ a timestamp and $k_\ccall$ a per-session digest
key for Layer 3. The handset then reads $\Reg[\org]$ and runs the checks described below. We write $\sid = \Hash(n \Vert \chid)$ for the
session identifier; it is unpredictable to anyone who has not seen the
nonce, which is what makes it usable as the key under which everything
else about the call is recorded.

The handset performs eleven ordered checks: credential presence and structure,
issuer resolution and claim matching, issuer signature, subject binding and holder
proof of possession, nonce freshness, channel authorisation, validity window and
revocation status. Cheap structural and registry failures are rejected first.

\paragraph{Grading the credential score.}
The outcome is a credential score
$\scred(\ccall,\org) \in \{0.0,\ 0.2,\ 0.5,\ 1.0\}$,
where $1.0$ means every check passed, $0.0$ means a check failed in a way
that can only be adversarial, $0.2$ means the credential is well formed
and correctly signed but outside its validity window, and $0.5$ means the
handset could not resolve revocation state and therefore does not know.

The grading is a diagnostic, not an input to a weighted sum. The gate
treats $\scred = 1.0$ as the only value satisfying its conjunct, so a
good audio score cannot promote a call with $\scred = 0.2$ or $0.5$, and
nothing whatsoever can promote $\scred = 0.0$. What the grading buys is
what the handset can honestly tell the recipient: at 4:12 she sees
``verified'', ``not verified'' with a reason, or ``blocked'', and those
are three different sentences.

One nuance in A3 is worth stating rather than glossing. Checks 6 and 7
defeat presentation of a stolen credential document, because the holder
must sign the challenge. They do not defeat theft of the organisation's
signing key, which is contained instead by revocation under check 11 and,
failing that, falls into the residual case A10 addresses.

\subsection{Layer 2: Optional, Session-Bound Path Attestation}
\label{sec:design:l2}

Let $Q \subseteq \{1,\dots,h\}$ be the participating subset of
intermediaries on the path. A participating hop $R_j$, on forwarding the
call, writes
\begin{equation}
\mathsf{attest}\bigl(\sid,\ j,\ \Hash(\Did_\org \Vert \chid_j \Vert
\sid),\ \tau_j\bigr)
\label{eq:attest}
\end{equation}
signed under $R_j$'s registered key, where $\chid_j$ is the hashed
calling identifier as $R_j$ observed it. The handset reads the
attestations for $\sid$ and forms the transcript $\mathcal{T}_\ccall =
\langle (j, \chid_j, \tau_j) \rangle_{j \in Q}$ ordered by $j$.

Including $\sid$ inside the attested hash is the whole construction, and
three properties follow. Attestations are not transferable, because
$\sid$ depends on a nonce this recipient generated for this call, so a
capture from an earlier call cannot be replayed into a later one.
Rewriting is detected wherever it is straddled: if an adversary rewrites
the identifier at a non-participating element between $R_j$ and
$R_{j'}$, then $\chid_j \neq \chid_{j'}$ and the transcript is
inconsistent, which needs two participants either side of the rewrite
rather than every hop participating. And non-participation is
distinguishable from tampering, so $\spath$ is ternary: $0$ if
$\mathcal{T}_\ccall$ is inconsistent or an attestation fails
verification, $\bot$ if $|Q| < 2$, and $1$ otherwise. We treat $\bot$ as
satisfying its conjunct in \Cref{eq:gate}, so a call over a
non-participating path is not blocked while an affirmatively inconsistent
one is. This is R5, and it is sound only because media substitution is
covered by Layer 3 rather than by the carriers.

Attestation is a write and therefore has a payer. A hop pays gas and
receives no direct benefit, which is the classic externality in transit
security. Our contract accepts attestations live, during the call, which
is what allows the handset to read the transcript during early media;
batching them after the session would cut the marginal cost but is
incompatible with a verdict inside the call, so we keep the live path and
report the measured per-hop cost explicitly in the evaluation.

\subsection{Layer 3: Digest Commitment and One Joint Anchor}

Call audio is legitimately altered in transit, so a cryptographic hash of
the waveform is useless. Following
AuthentiCall~\cite{reaves2017authenticall} and the fingerprinting it
draws on~\cite{haitsma2002robust,cano2005review}, we use a robust digest
$\Dig(\cdot)$ mapping a short speech frame to a bit string that preserves
perceptual content while tolerating transcoding and loss. The originating
endpoint transmits $d_i = \Dig(a_i)$ over the side channel keyed under
$k_\ccall$; the handset computes $d_i'$ from the audio it received and
accepts frame $i$ when $\ber(d_i,d_i') \leq \tau_{\ber}$, with $\sint$
the fraction of accepted frames over a sliding window.

Nothing so far needs a ledger. At teardown the originating endpoint
computes the Merkle root $\Root_\ccall$ over $(d_1,\dots,d_T)$ and the
path commitment $\Ptr_\ccall = \Hash(\mathcal{T}_\ccall)$, and submits
one transaction $\mathsf{anchor}(\Did_\org, \sid, \Root_\ccall,
\Ptr_\ccall, T, |Q|)$. Three consequences follow. No audio and no digest
reaches the chain, only roots. The mandatory anchor cost is constant in
duration and hop count, since a ten-second and a fifty-minute call anchor
one root each, and a two-hop and a twelve-hop path one transcript
commitment each. Optional Layer~2 attestations remain one write per
participating hop, so total system writes are $1+|Q|$; both costs are
reported in \Cref{sec:eval:throughput}. And the evidence survives
the call held by nobody: given a retained recording, any third party can
recompute $d_i'$, open the Merkle path and check it against a root the
organisation signed and the chain timestamped, so the recipient cannot
fabricate a favourable root and the organisation cannot repudiate one it
signed. This is R6, and its evidential value is conditional on a
recording being retained, which is a property of the design rather than
an oversight.

\subsection{Layer 4 and the Composition Rule}
\label{sec:design:gate}

The liveness component returns $\sbeh(\ccall) \in [0,1]$, an estimate
that the audio in the first $W$ seconds is natural rather than synthetic,
computed locally from a cepstral front end with a neural back end of the
kind surveyed in the spoofing detection
literature~\cite{tak2022wav2vec}. The choice of detector is not a
contribution and the argument below holds for any detector with
comparable error characteristics. What the layer is for is narrow: not a
second opinion on integrity, which Layer 3 settles cryptographically, nor
on route, which Layer 2 settles where it applies, but solely A10, the
case in which the registered endpoint itself emitted synthetic audio
under a valid key, where every digest and attestation therefore verifies.

Writing $\mathsf{verified}(\ccall,\org)$ for the event that the handset
presents the call as verified, the policy is
\begin{equation}
\mathsf{verified}(\ccall,\org) \iff
\underbrace{\scred = 1}_{\text{L1}} \wedge
\underbrace{\spath \neq 0}_{\text{L2}} \wedge
\underbrace{\sint \geq \theta_{\mathrm{int}}}_{\text{L3}} \wedge
\underbrace{\sbeh \geq \theta}_{\text{L4}}.
\label{eq:gate}
\end{equation}
The first three conjuncts are necessary and their thresholds come from
cryptographic and signal-processing analysis; the fourth may only remove
a verdict the first three granted. This is not a design preference but
the only non-degenerate option available in the linear family.

\begin{proposition}[Gating is canonical in positive linear fusion]
\label{prop:gating}
For a positive-weight linear fusion of binary cryptographic conjuncts and
a behavioural score, any non-degenerate rule that prevents behavioural
evidence from rescuing a failed cryptographic conjunct reduces to
\Cref{eq:gate} for some behavioural threshold $\theta$.
\end{proposition}
The full parameterisation and proof are given in \Cref{app:proof}. The
consequence is that the cryptographic layers are conjunctive rather than
tradeable: their individual fusion weights carry no safe notion of
relative importance, and $\theta$ is the only surviving operating
parameter. We therefore treat $\spath=\bot$ as satisfying the path
conjunct; otherwise every non-participating path would be refused.

Because compromise of a registered key is rare relative to legitimate
credentialled traffic, the prior that a call passing all three
cryptographic conjuncts is an A10 attack is small, and $\theta$ must
therefore sit at a low false demotion rate rather than near the equal
error rate at which such detectors are usually reported. Detectors of
this family lose most of their discrimination under corpus
shift~\cite{muller2022generalize,barrington2024deepspeak}, so we fix
$\theta$ from a target false demotion rate on held-out genuine data
rather than from equal error rate. We report no detector results in this
paper and make no claim about the residual protection Layer 4 provides
against A10; what we claim is architectural, that the weakness of the
detector cannot become a weakness of the guarantee.

  \begingroup
  \let\bvisoriginallabel\label
  \renewcommand{\label}[1]{\bvisoriginallabel{fig:seq}}  
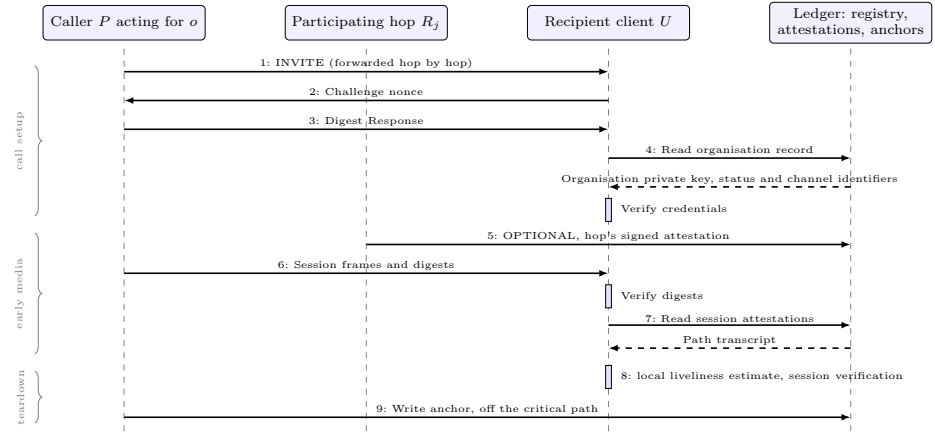
\begin{figure}[t]
\centering
\resizebox{\textwidth}{!}{%
\begin{tikzpicture}[
  font=\scriptsize,
  head/.style={draw, rounded corners=2pt, align=center, inner sep=3pt,
               minimum height=6mm, text width=26mm, fill=blue!6},
  msg/.style={-{Latex[length=1.6mm]}, thick},
  ret/.style={-{Latex[length=1.6mm]}, thick, dashed},
  lbl/.style={font=\tiny, midway, above, inner sep=1.2pt},
  act/.style={draw, fill=blue!12, minimum width=1.6mm, inner sep=0pt}
]

\def\xP{0}   \def\xR{4.2}   \def\xU{8.4}   \def\xL{12.6}
\def\ybot{-7.1}

\node[head] (P) at (\xP,0) {Caller $P$ acting for $\org$};
\node[head] (R) at (\xR,0) {Participating hop $R_j$};
\node[head] (U) at (\xU,0) {Recipient client $U$};
\node[head] (L) at (\xL,0) {Ledger: registry, attestations, anchors};

\foreach \x in {\xP,\xR,\xU,\xL}
  \draw[dashed, gray] (\x,-0.45) -- (\x,\ybot);

\draw[decorate, decoration={brace, amplitude=3.5pt}, gray]
  (-1.55,-0.75) -- (-1.55,-3.35)
  node[midway, left=2pt, font=\tiny, rotate=90, anchor=south] {call setup};
\draw[decorate, decoration={brace, amplitude=3.5pt}, gray]
  (-1.55,-3.65) -- (-1.55,-5.75)
  node[midway, left=2pt, font=\tiny, rotate=90, anchor=south] {early media};
\draw[decorate, decoration={brace, amplitude=3.5pt}, gray]
  (-1.55,-6.05) -- (-1.55,-6.95)
  node[midway, left=2pt, font=\tiny, rotate=90, anchor=south] {teardown};

\draw[msg] (\xP,-0.85) -- (\xU,-0.85) node[lbl] {1: INVITE (forwarded hop by hop)};
\draw[msg] (\xU,-1.35) -- (\xP,-1.35) node[lbl] {2: Challenge nonce};
\draw[msg] (\xP,-1.85) -- (\xU,-1.85)
  node[lbl] {3: Digest Response};
\draw[msg] (\xU,-2.35) -- (\xL,-2.35) node[lbl] {4: Read organisation record};
\draw[ret] (\xL,-2.85) -- (\xU,-2.85) node[lbl] {Organisation private key, status and channel identifiers};
\node[right=1mm, font=\tiny] at (\xU,-3.25) {Verify credentials};
\draw[act] (\xU-0.05,-3.05) rectangle (\xU+0.05,-3.45);

\draw[msg] (\xR,-3.85) -- (\xL,-3.85)
  node[lbl] {5: OPTIONAL, hop's signed attestation};

\draw[msg] (\xP,-4.35) -- (\xU,-4.35)
  node[lbl] {6: Session frames and digests};
\node[right=1mm, font=\tiny] at (\xU,-4.75) {Verify digests};
\draw[act] (\xU-0.05,-4.55) rectangle (\xU+0.05,-4.95);

\draw[msg] (\xU,-5.25) -- (\xL,-5.25) node[lbl] {7: Read session attestations};
\draw[ret] (\xL,-5.65) -- (\xU,-5.65)
  node[lbl] {Path transcript};
\node[right=1mm, font=\tiny] at (\xU,-6.15)
  {8: local liveliness estimate, session verification};
\draw[act] (\xU-0.05,-5.95) rectangle (\xU+0.05,-6.35);

\draw[msg] (\xP,-6.85) -- (\xL,-6.85)
  node[lbl] {9: Write anchor, off the critical path};

\end{tikzpicture}}
\caption{Sequence of one verified session. Steps 1 to 4 complete within
call setup and cost the recipient reads only. Step 5 is contributed
independently by each participating hop and is simply absent for hops
outside $Q$, which yields $\spath=\bot$ rather than a failure. Steps 6 to
8 run inside the early media window. Step 9 is the single anchoring
write, made by the originating endpoint at teardown and therefore off the
path the recipient waits on.}
\label{fig:seq}
\end{figure}

  \endgroup

\section{Implementation}
\label{sec:impl}

\subsection{Contract}

The registry is implemented in Solidity against the Ethereum virtual
machine~\cite{wood2014yellow}, compiled with solc 0.8.24 and the
optimiser enabled at 200 runs. Hop attestations are append-only and keyed
by $(\sid, j)$, so a verifier retrieves a transcript with one indexed
read and no intermediary can overwrite another's entry. Session anchors
are write-once and keyed by $(\Did_\org, \sid)$, which makes an anchor
addressable by a party holding the session context without that key
revealing the channel to a casual observer. Administrative operations are
guarded by a registrar role, and every state-changing operation emits an
event so verifiers maintaining a local index can follow the contract
without replaying full state. Key rotation retains superseded keys with
their validity intervals rather than overwriting, so an anchor written
before a rotation is still verifiable against the key that signed it;
without this, routine key management would silently destroy the
auditability of R6. Revocation is a state transition, not a deletion.
The implementation follows the registry, attestation and anchor interfaces
described above; the source artefact contains the complete contract interface.

\subsection{Verifier Client}

The handset performs all four layers locally, and nothing about the call
leaves the device as a condition of verification. Resolved records are
cached with a short time to live, and an entry is invalidated immediately
on observing a status change event, which is what gives revocation its
practical effect. Attestation reads are session-scoped and not cached.
The digest engine is kept separate from the detector, so that no code
path allows a behavioural score to reach either the digest comparison or
the chain.

\subsection{What Is Implemented, and What Is Not}

The prototype implements the registry and registrar controls,
session-bound path attestation, session anchoring, the digest pipeline and
the composition gate. Layer~4 is deliberately detector-agnostic: the
architecture consumes a local behavioural score, but this paper does not
claim accuracy for a particular synthetic-speech detector.

The scenario suite contains exactly 27 Hardhat tests: 26 assertions
grouped under A1--A10 plus one coverage-summary consistency test. Each test runs from an isolated Hardhat fixture state containing a
deployed \texttt{BVIRegistry}; all 27 pass. The scope of those tests follows
the architecture. For A9 they
establish that the digest commitment is session-bound, joint with the
path commitment and immutable once written, while media separability is
measured separately. For A10 they establish the gating property that no
detector score can enter chain state or promote a verdict; detector miss
and false-demotion rates are not evaluated here.

Production carrier deployment is not implemented. Path results are
simulated over a modelled path, and public-network confirmation latency
is not measured. We keep measured, simulated and not-yet-done results
separate throughout.

\section{Evaluation}
\label{sec:eval}

We report measured, simulated and derived results separately. Gas figures
come from a local Hardhat chain with solc 0.8.24, the optimiser at 200
runs and EVM version \texttt{paris}; each operation value is the median
of five transaction receipts. The path study is one seeded simulation
(seed 20260909) with 20{,}000 sessions per cell. No recipient-side timing
experiment is claimed in this paper.

\subsection{Attack Coverage and Automated Tests}
\label{sec:eval:coverage}

\begin{table}[tb]
\centering
\caption{Attack coverage by mechanism. $\checkmark$ denotes coverage,
$\circ$ partial coverage and -- no coverage. The 27 A1--A10 tests exercise
BVI's contract-side behaviour; A9/A10 scope is stated in the text.}
\label{tab:coverage}
\footnotesize
\setlength{\tabcolsep}{4.2pt}
\begin{tabular}{@{}llcccc@{}}
\toprule
ID & Attack & BBCA & Digest & Detector & BVI \\
\midrule
A1  & No credential          & -- & -- & -- & $\checkmark$ \\
A2  & Self-signed credential & -- & -- & -- & $\checkmark$ \\
A3  & Stolen/compromised credential & $\circ$ & -- & -- & $\checkmark^{\dagger}$ \\
A4  & Tampered claim         & $\circ$ & -- & -- & $\checkmark$ \\
A5  & Revoked credential     & -- & -- & -- & $\checkmark$ \\
A6  & Stale key epoch        & -- & -- & -- & $\checkmark$ \\
A7  & Replayed assertion     & -- & $\checkmark$ & -- & $\checkmark$ \\
A8  & In-path rewrite        & $\circ$ & -- & -- & $\checkmark$ \\
A9  & Media substitution     & -- & $\checkmark$ & -- & $\checkmark$ \\
A10 & Synthetic speech       & -- & -- & $\checkmark$ & $\checkmark^{\ddagger}$ \\
\bottomrule
\end{tabular}
\end{table}

The matrix explains the composition: A8 leaves audio unchanged, A9 can
leave route and speaker type unchanged, and A10 can originate from a
credentialled endpoint, so no single mechanism subsumes the others. The
A3 mark combines verifier-side holder binding/proof of possession with
revocation containment for signing-key compromise; the pre-revocation
window remains a stated limitation. A10 coverage is architectural only:
the local detector may demote, but its miss rate is not measured here.

The 27-test suite contains 2, 3, 3, 2, 2, 2, 3, 3, 3 and 3 tests for
A1--A10 plus one coverage-summary check. For A9/A10 the contract tests
cover commitment/gating properties, not media or detector accuracy. Case
U remains an explicit indeterminate outcome.

\subsection{Path Evidence Under Partial Participation}
\label{sec:eval:path}

The unified path study uses 20{,}000 sessions per cell for 3, 5, 8 and
12 hops, independent participation and a 30 per cent tamper rate.
At five hops and $p=0.25$, BVI detects 20.6 per cent of randomly located
rewrites versus 0.1 per cent for the every-hop baseline; at least one
path observation exists in 75.7 per cent of sessions, while insufficient
evidence returns $\bot$ rather than failure. The anchor itself is
independent of path participation.

The path-length advantage is average-case. A strategic attacker who knows
participation reduces detection to approximately $p^2$; at $p=0.5$ and
12 hops, average and strategic BVI detection are 82.8 and 24.6 per cent,
while the every-hop baseline is below 0.1 per cent. Five-hop false
demotion is 3.8 per cent at $p=0.5$ and 6.4 per cent at full
participation, driven by legitimate rewrites at unregistered border
controllers. These are simulated results.

\begin{table}[tb]
\centering
\caption{Five-hop path results, 20{,}000 sessions per cell, seed
20260909. Detection is over tampered sessions; ``strategic'' knows the
participating set.}
\label{tab:participation}
\footnotesize
\setlength{\tabcolsep}{4.2pt}
\begin{tabular}{@{}rrrrrr@{}}
\toprule
$p$ & BVI avg. & BVI strategic & BBCA & Path evidence & BBCA record \\
\midrule
0.00 & 0.0\%   & 0.0\%   & 0.0\%   & 0.0\%   & 0.0\% \\
0.25 & 20.6\%  & 5.9\%   & 0.1\%   & 75.7\%  & 0.1\% \\
0.50 & 56.6\%  & 24.8\%  & 3.1\%   & 96.9\%  & 3.1\% \\
0.75 & 82.7\%  & 55.7\%  & 23.5\%  & 99.9\%  & 23.9\% \\
1.00 & 100.0\% & 100.0\% & 100.0\% & 100.0\% & 100.0\% \\
\bottomrule
\end{tabular}
\end{table}

\subsection{Content Layer}

At 17 Bark-spaced bands and a 20~ms hop the digest stream is
800~bits/s. In controlled fully voiced synthetic sessions, a fixed benign
threshold gives 100 per cent detection at 1 and 5 per cent FPR for all
tested splice fractions (2--100 per cent), with local splice BER
0.479--0.499 and a shortest analysis window of about 240~ms. These are
upper-bound feasibility results; real speech and gateway carriage remain
outside this experiment.

\subsection{On-Chain Cost and the Throughput Ceiling}
\label{sec:eval:throughput}

The median \texttt{anchorSession} cost is 91{,}547 gas; across 30
duration-by-hop combinations it ranges only from 91{,}523 to 91{,}547,
confirming fixed-width cost. Each participating intermediary adds one
optional 116{,}392-gas attestation, so total BVI writes are $1+|Q|$;
under full participation BBCA requires 2, 5, 8, 14 and 23 writes for
$h=1,2,3,5,8$.

A representative 60M-gas, two-second rollup yields about 328 anchors/s
on a BVI-dedicated chain. Batching is bounded near 1.30$\times$ because
per-anchor storage writes remain, leaving higher capacity or horizontal
partitioning as the main scaling levers. Public-network confirmation
latency is not measured and is off the recipient's call path.

\section{Discussion and Conclusion}
\label{sec:disc}
\label{sec:conclusion}

The main limits are explicit. The splice corpus is fully voiced and
synthetic; path evidence is simulated under a specified participation
model; Layer~4 detector accuracy is unevaluated; and the 27 scenario
tests establish contract-side behaviour rather than real carrier
deployment. Organisation registration remains a bootstrap requirement,
recipient interpretation is untested, and a still-valid signing key or
human caller creates a pre-revocation exposure window that the ledger can
audit but cannot retroactively prevent.

Within those bounds, BVI verifies a live, session-bound organisational
credential on the arriving channel without a call-time chain write. Its 10-by-4
coverage matrix and 27 passing scenario tests exercise the composed design.
The roughly 91.5k-gas anchor keeps session state fixed-width, while path evidence
improves with carrier participation.

\bibliographystyle{splncs04}
\bibliography{references}


\appendix
\setcounter{section}{0}
\renewcommand{\thesection}{\Alph{section}}
\renewcommand*{\theHsection}{appendix.\Alph{section}}

\section{Proof of \Cref{prop:gating}}
\label{app:proof}
Let $c_1,\dots,c_k\in\{0,1\}$ be cryptographic conjunct outcomes and
$b\in[0,1]$ a behavioural score. Let
$S=\sum_j\lambda_jc_j+\lambda_b b$, with $\lambda_j>0$, $\lambda_b>0$,
$\Lambda=\sum_j\lambda_j$ and $\Lambda+\lambda_b=1$, and accept iff
$S\geq\tau$. If $c_j=0$ and all other conjuncts equal one, acceptance is
possible for some $b$ iff $\tau\leq1-\lambda_j$; avoiding this for every
$j$ requires $\tau>1-\min_j\lambda_j$. With all $c_j=1$, the rule accepts
iff $b\geq(\tau-\Lambda)/\lambda_b$. The behavioural term is therefore
inert when $\tau\leq\Lambda$ and otherwise acts only as the threshold
$\theta=(\tau-\Lambda)/\lambda_b$. Hence every non-degenerate positive
linear rule that prevents behavioural evidence rescuing a failed
cryptographic conjunct reduces to \Cref{eq:gate}. \qed

\section{Reproducibility}
\label{app:repro}
\paragraph{Scenario tests.} The Hardhat suite contains 27 tests:
2/3/3/2/2/2/3/3/3/3 under A1--A10 plus one coverage-summary check.
The fixture deploys \texttt{BVIRegistry}, and Hardhat restores an isolated
fixture state for each test. For A9 the suite
tests the contract-side commitment properties; for A10 it tests the
demotion-only gating boundary rather than detector accuracy.

\paragraph{Gas.} Measurements use a local Hardhat chain, solc 0.8.24,
EVM version \texttt{paris}, and the optimiser at 200 runs; reported
operation costs are medians of five transaction receipts. The
fixed-width anchor sweep covers 30 duration-by-hop combinations and
ranges from 91{,}523 to 91{,}547 gas.

\paragraph{Path simulation.} The reported path figures use seed 20260909
and 20{,}000 sessions per cell for path lengths 3, 5, 8 and 12.
Participation is drawn independently per hop and tampering independently
per session with probability 0.3. The average-case attacker places the
rewrite uniformly; the strategic attacker observes participation and
chooses the tamper point accordingly.

\paragraph{Digest.} The controlled content experiment uses 17 Bark-spaced
bands, a 20~ms hop, 800~bits/s and 400-frame synthetic sessions. The
threshold is derived from held-out benign data and held fixed across
splice fractions.

\paragraph{Attack construction.} A1 omits the credential; A2 substitutes
a self-issued key; A3 covers stolen-holder material and the
compromise/revocation branch; A4 tampers with a claim; A5 uses revoked
state; A6 a stale key epoch; A7 replays a presentation into a fresh
session; A8 rewrites the observed identifier; A9 substitutes
authenticated media; and A10 preserves the cryptographic conjuncts while
supplying synthetic speech.

\end{document}